\documentclass[fleqn,usenatbib]{mnras}
\usepackage{newtxtext,newtxmath}
\usepackage[T1]{fontenc}
\usepackage{color,soul}
\DeclareRobustCommand{\VAN}[3]{#2}
\let\VANthebibliography\thebibliography
\def\thebibliography{\DeclareRobustCommand{\VAN}[3]{##3}\VANthebibliography}

\usepackage{graphicx}	
\usepackage{amsmath}	

\title[Individual element isochrone sensitivity]{Individual element sensitivity for stellar evolutionary isochrones}

\author[Worthey et al.]{G. Worthey,$^{1}$\thanks{E-mail: gworthey@wsu.edu (GW)}
Xiang Shi,$^{1}$
Tathagata Pal,$^{1}$
H.-c. Lee,$^{2}$
and B. Tang$^{3}$
\\
$^{1}$Department of Physics and Astronomy, Washington State University, Pullman, WA 99164-2814, USA\\
$^{2}$Department of Physics and Astronomy, The University of Texas Rio Grande Valley, 1201 W. University Dr., Edinburg, TX 78539, USA\\
$^{3}$School of Physics and Astronomy, Sun Yat-sen University, Zhuhai 519082, People's Republic of China\\
}

\date{Accepted 2022 Jan 26. Received 2021 Nov 11.}

\pubyear{2022}

\begin{document}
\label{firstpage}
\pagerange{\pageref{firstpage}--\pageref{lastpage}}
\maketitle

\begin{abstract}
Stellar evolution calculations with variable abundance ratios were used to gauge the effects on temperatures, luminosities, and lifetimes in various phases. Individual elements C, N, O, Mg, Si, and Fe were included. Most of the effect relevant to integrated light models is contained in the temperature variable, as opposed to the timescale or luminosity. We derive a recipe for including abundance-sensitive temperature effects that is applicable to existing isochrone grids. The resultant enhanced isochrones are incorporated into composite stellar population models and compared with galaxy data from the Sloan Digital Sky Survey. A severe oxygen-age degeneracy is apparent, 2 - 3 Gyr per 0.1 dex in [O/R], where R represents a heavy element such as Fe. Over the range of early-type galaxy velocity dispersion, the spans of all abundance ratios are reduced but the age range increases, systematically older. Allowing Fe-peak elements the freedom to vary accentuates this increase of age span. Overall, these results sharpen the age-mass correlation known as downsizing but decrease the steepness of abundance ratio gradients. Both of these observations, in turn, imply a more robust contribution from gas free mergers in the histories of typical elliptical galaxies. 
\end{abstract}




\section{Introduction}
Galaxies loom large in the quest to understand the construction of our universe. These ubiquitous units of matter, however, are complicated in structure and history. Furthermore, the integrated light spectrum observed from a galaxy is often the only information we can obtain. Comparison of observed spectra to model spectra is one fruitful way to approach studies of chemical abundances, age structures, and kinematics of these systems.  
In principle, a spectrum can be obtained for a galaxy at any redshift, so such techniques are broadly applicable.

Most models today use a single-stellar-population (SSP) to give an estimate of a galaxy's mean age \citep{2003MNRAS.344.1000B,2003MNRAS.339..897T,Worthey_2014,2015MNRAS.449.1177V}. The idea that a galaxy originates at one time and then passively evolves after that is no longer considered realistic, but the SSP models continue to be useful when used correctly. For example, early-type galaxies show evidence for later star formation  \citep{10.1111/j.1365-2966.2008.13132.x} independent of evidence from redshift surveys  \citep{2006ApJ...651..120B,2004ApJ...608..752B,2007ApJ...665..265F} that also show a need for later mass assembly.

In addition, spectra carry chemical abundance information. For example, very massive galaxies appear to have enhanced [Mg/Fe] \citep{1989PhDT.......149P,1994A&A...288...57M}. Interestingly, the first paper to model the situation \citep{1992ApJ...398...69W} also mentioned that isochrone-based models were inadequate to truly measure a rigorous [Mg/Fe] ratio. They recognized that the addition of Mg into a chemical mixture would change stellar opacities and therefore might alter stellar temperatures. Specifically, increasing Mg (alone, not in solar mixture lockstep) might make giant stars cooler, leading to increased metallic line strengths in general, including Mg features.

Early literature on this topic [c.f. historical review in $\S$2 of \cite{1998PASP..110..888W}] implied that [Mg/Fe]>0 for massive elliptical galaxies, but not lower-mass galaxies around a Milky Way mass.  The possible processes to develop this abundance trend include \citep{1992ApJ...398...69W,2000AJ....120..165T} star formation time scales, variable initial mass function (IMF), variable binary fraction, and selective loss mechanisms such as supernova (SN) driven winds. All explanations hinge on the separation of supernova flavors. Type Ia supernovae produce iron peak elements but require a population of white dwarfs to develop before exploding. They take longer to occur and may involve lower-mass progenitors. Type II SNs explode almost immediately after formation and yield a broader spectrum of elements, though the exact yields vary with progenitor mass and metallicity \citep{1995ApJS..101..181W,2006NuPhA.777..424N}.  \cite{1994A&A...288...57M} investigated the timescale hypothesis to infer a formation time for the most massive ellipticals of $\sim 3\times 10^8$ years based on Type Ia versus Type II supernova timescales. 

We prefer the term "light elements" to "alpha elements" for this work. The term alpha elements refers to nuclei formed by adding alpha particles ($^4_2$He) to $^{12}_{\ 6}$C, though carbon itself is generally not included amongst the alpha capture elements. The list is O, Ne, Mg, Si, S, Ar, Ca, and Ti, and most of these elements (not C) are enhanced in metal-poor Milky Way halo stars relative to Fe \citep{2017MNRAS.465.1586F}. Fewer of them are enhanced in massive elliptical galaxies or the Milky Way bulge \citep{2021ApJ...909...77G,2014MNRAS.445.1538T,1998PASP..110..888W}. Using isochrones whose temperatures remained fixed as element ratios were varied, \cite{Worthey_2014} found that C, N, O, Na, Mg, and Si have elevated abundances, but Ca and Ti do not (with no information for the noble gases). The SN hypothesis may still hold valid as a cause if the origins of Ca and Ti are Type Ia SN or lower-mass Type II progenitors. 
However, besides the Type Ia/Type II ratio, the composition and mass of the supernova's progenitor has been linked to its effects on the chemical abundance of the galaxy in whole \citep{Tang_2015,1995ApJS..101..181W,2013ARA&A..51..457N,2012ApJ...760L..38T,2006NuPhA.777..424N}.

\citet{2007ApJ...666..403D} discussed the impact of individual element sensitivity on stellar model evolution. The common effect being an increase in the abundance of elements C, N, and O result in an increased effective temperature and a decreased stellar lifetime. The effective temperatures of stars was shown to decrease when Mg, Si, and Fe abundances were increased but differ on their effects on stellar lifetime. The isochrones used considered temperature shifts over a constant heavy element mass fraction ($Z$). Our dynamic isochrones of this paper will compensate for temperature shifts based on detailed abundance changes. Stars at lower metallicities were discussed by \citet{2012ApJ...755...15V} with broadly similar results after accounting for the fact that their treatment was at constant [Fe/H] rather than constant $Z$.

Extending the work of \citet{2007ApJ...666..403D} in metallicity span and mass range, we hope in this work to systematize the effects of individual element abundance changes enough to incorporate into integrated-light models. We then apply the new models to galaxy spectra to assess what previous astrophysical conclusions might be at risk.

The remainder of the paper is organized as follows: Section 2 discusses the modeling effort, starting from stellar evolution modeling and proceeding through integrated light models for stellar populations. Results obtained with comparisons to Sloan Digital Sky Survey (SDSS) average spectra are presented in $\S$3. A discussion follows in $\S$4 and concluding remarks appear in $\S$5. 


\section{Calculations}

\subsection{Inclusion of altered abundance mixtures}

Suites of stellar evolutionary tracks with altered abundances were calculated with MESA \citep{2011ApJS..192....3P} version 4942. The calculations sampled heavy element mass fractions $0.0001 < Z < 0.04$ at two values of $Y$ and for masses $0.06 < m / M_\odot < 3$ (see Table \ref{tab:zy}). \cite{1998SSRv...85..161G} abundances were used, along with the custom abundance-sensitive opacity tables calculated for \cite{2007ApJ...666..403D}. CNO-sensitive "Type 2" opacities were employed within MESA. For a grid beyond scaled-solar, one element was increased, and the rest of the abundances were rescaled so that heavy element mass fraction $Z$, along with $X$ and $Y$ remained constant. The surface boundary condition was set by Eddington gray atmospheres that responded to the abundance changes, and the nuclear reaction network was altered to accommodate the new abundance patterns as well. Six separate elements were considered in this way: C, N, O, Mg, Si, and Fe. Evolution past the helium flash was generally deemed unnecessary for this experiment. 

The abundance changes affect evolutionary timescale, luminosity, and surface temperature for each star. However, the timescales and luminosities generally oppose each other according to the fuel consumption theorem \citep{1986ASSL..122..195R,1989ApJS...71..817B,1998MNRAS.300..872M} to yield nearly the same energy output. For purposes of integrated light, we expect that stellar temperature changes result in the largest overt spectral changes when integrated.

We fit the tracks using a multivariate polynomial regression based on the Levenberg-Marquardt algorithm \citep{Levenberg,Marquardt}. We sought solutions of the form 

\begin{equation}
    dT(X) = \frac{d\rm{[X/R]}}{O(X) } f(X, g, {\rm [M/H]}) \, ,
    \label{eq:fit}
\end{equation}
where $dT$ is the change in (ten-based) log of surface temperature in log degrees Kelvin and $g$ is log $g$ in the traditional cgs units. [M/H] $\equiv $ log $[N_{\rm heavy} / N_{\rm H}] - $ log $[N_{{\rm heavy}},\odot / N_{{\rm H},\odot}]$ $\approx $ log $Z/Z_\odot$. The quantity $R$ stands for ``generic heavy element,'' employed because we can change Fe abundance by itself, so the normal notation ([Fe/Fe]) would become paradoxical compared to [Fe/R]. The quantity d[X/R] is the logarithmic change in abundance for element $X$, and [R/H] equates to [Fe/H] until Fe changes by itself. The function $O$ is a nod to the increments used in the evolutionary calculations, and usually has the value 0.3 dex, a factor of two. However, for carbon, we wished to avoid accidentally manufacturing carbon stars, and 0.3 dex would make the C abundance exceed the O abundance, so we used a decrement of 0.2 dex for carbon. See Table \ref{tab:O}. Heavy element mass fraction $Z$ varies for the fitting process, but each d[X/R] is evaluated at a fixed $Z$.

\begin{table}
	\centering
	\caption{Grid mass fractions, two $Y$ values per $Z$. }
	\label{tab:zy}
	\begin{tabular}{lcc} 
		\hline
		$Z$ & $Y$ & $Y$ \\
		\hline
		0.0001 & 0.21 & 0.27 \\
	    0.0003 & 0.22 & 0.27 \\
		0.001  & 0.22 & 0.27 \\
		0.003  & 0.22 & 0.27 \\
		0.01   & 0.25 & 0.30 \\
		0.02   & 0.25 & 0.30 \\
		0.04   & 0.30 & 0.35 \\
		\hline
	\end{tabular}
\end{table}

\begin{table}
	\centering
	\caption{Values for the $O$ function in equation \ref{eq:fit}. }
	\label{tab:O}
	\begin{tabular}{lc} 
		\hline
		Element $X$ & $O(X)$ \\
		\hline
		C & 0.2 \\
		N & 0.3 \\
		O & 0.3 \\
	    Mg & 0.3 \\
		Si & 0.3 \\
		Fe & 0.3 \\
		\hline
	\end{tabular}
\end{table}

\begin{figure}
	\includegraphics[width=\columnwidth]{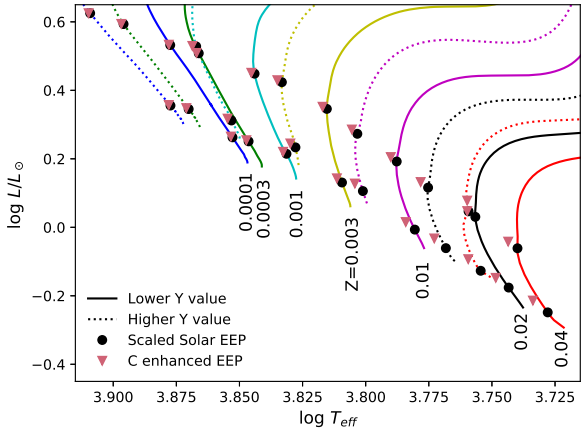}  
    \caption{Main sequence evolution in the model grid for some 1 M$_\odot$ models showing the effects of helium and carbon enhancements. The equivalent evolutionary points (EEPs) illustrated are defined by two values of central hydrogen mass fraction ($X = 0.6$ and $X = 0.1$). }
    \label{fig:eep_overview}
\end{figure}

\begin{figure}
	\includegraphics[width=\columnwidth]{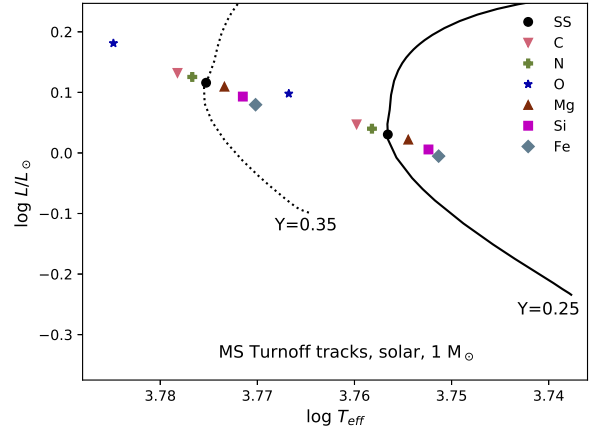}  
    \caption{The effects of increasing individual elements at constant [R/H] in the H--R diagram near the main sequence turnoff. Tracks at solar composition and one solar mass but two $Y$ values are shown. At the EEP defined by central $X=0.1$, the effects of individual element enhancements are shown. Increasing the lighter elements (C, N, O) heat the main sequence while heavier ones (Mg, Si, Fe) cool it.}
    \label{fig:eep_ms}
\end{figure}

\begin{figure}
	\includegraphics[width=\columnwidth]{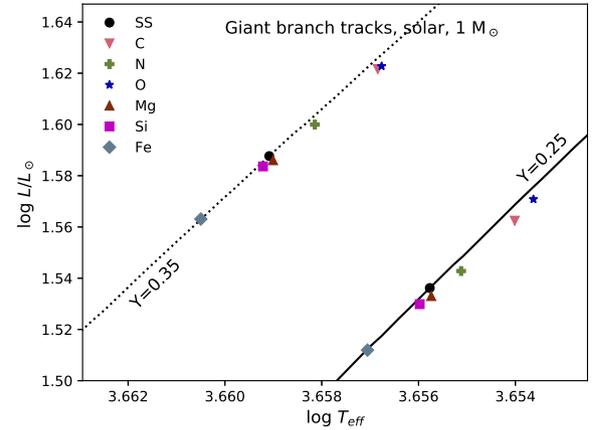}  
    \caption{The effects of increasing individual elements at constant [R/H] on the RGB. Tracks at solar composition and one solar mass but two $Y$ values are shown. At the EEP defined by helium core mass reaching $M_C/M=0.25$, the effects of individual element enhancements are shown. The spread of points parallel to the giant branch is due to timing differences inherited from the main sequence more than actual shifts of stellar structure.}
    \label{fig:eep_rgb}
\end{figure}

Initially, we envisioned the use of equivalent evolutionary points (EEPs) \citep{1987ryil.book.....G} as invariate anchors from which we could compute $\partial T / \partial X$ derivatives, and this method works well enough in the main sequence region (cf. Figs.\ref{fig:eep_overview} and \ref{fig:eep_ms}). However, using EEPs on the red giant branch (RGB) appears to generate misleading results. As shown in Figure \ref{fig:eep_rgb}, the star's previous history affects its nuclear clock and internal structure such that points spread primarily along the RGB rather than horizontally in the temperature direction. For the RGB, therefore, we held luminosity constant rather than EEP. This generates a discontinuity among the subgiants where neither EEP nor luminosity is an acceptable invariate. In practice, we fit the RGB and MS separately. During the fitting process, we treated nonlinear numerical artifacts as noise and judiciously deleted discrepant points. The resultant multivariate polynomials are listed in Tables \ref{tab:f} and \ref{tab:f2}.

\begin{table}
	\centering
	\caption{List of the $f$ functions in equation \ref{eq:fit} for dwarfs. }
	\label{tab:f}
	\begin{tabular}{ll} 
		\hline
		$X$ & $f(X,g,{\rm [M/H]})$ \\
		\hline
C   & $ -0.455918\times 10^{+00} +0.324351\times 10^{-01} {\rm [M/H]} +0.219075\times 10^{+00} g $ \\
    & $ -0.262079\times 10^{-01} g^2 -0.735549\times 10^{-02}{\rm [M/H]} $ \\
N   & $ -0.832092\times 10^{-01} +0.105843\times 10^{-01} {\rm [M/H]} +0.418383\times 10^{-01} g $ \\
    & $ -0.516371\times 10^{-02} g^2 -0.231520\times 10^{-02}{\rm [M/H]} $ \\
O   & $ -0.649515\times 10^{+00} +0.133922\times 10^{+00} {\rm [M/H]} +0.324449\times 10^{+00} g $ \\
    & $ -0.391601\times 10^{-01} g^2 -0.276506\times 10^{-01}{\rm [M/H]} $ \\
Mg  & $ +0.341413\times 10^{-01} -0.328280\times 10^{-02} {\rm [M/H]} -0.172424\times 10^{-01} g $ \\
    & $ +0.207316\times 10^{-02} g^2 +0.555231\times 10^{-03}{\rm [M/H]} $ \\
Si  & $ -0.154322\times 10^{-01} -0.241229\times 10^{-02} {\rm [M/H]} +0.237230\times 10^{-02} g $ \\
Fe  & $ +0.152259\times 10^{-01} -0.192085\times 10^{-01} {\rm [M/H]} -0.150910\times 10^{-01} g $ \\
    & $ +0.229713\times 10^{-02} g^2 +0.361368\times 10^{-02}{\rm [M/H]} $ \\
		\hline
		
	\end{tabular}
\end{table}

\begin{table}
	\centering
	\caption{List of the $f$ functions in equation \ref{eq:fit} for giants. }
	\label{tab:f2}
	\begin{tabular}{ll} 
		\hline
		$X$ & $f(X,g,{\rm [M/H]})$ \\
		\hline
C   & $ -0.541111\times 10^{-03} -0.768735\times 10^{-04} {\rm [M/H]} +0.381417\times 10^{-03} g $ \\
    & $ -0.908027\times 10^{-04} g^2 +0.140252\times 10^{-04}{\rm [M/H]} g $ \\
N   & $ -0.274718\times 10^{-03} -0.248253\times 10^{-04} {\rm [M/H]} +0.229881\times 10^{-03} g $ \\
    & $ -0.582940\times 10^{-04} g^2 +0.554453\times 10^{-05}{\rm [M/H]} g $ \\
O   & $ -0.510599\times 10^{-03} -0.761868\times 10^{-04} {\rm [M/H]} +0.411131\times 10^{-03} g $ \\
    & $ -0.943409\times 10^{-04} g^2 +0.281746\times 10^{-04}{\rm [M/H]} g $ \\
Mg  & $ -0.862162\times 10^{-04} +0.127068\times 10^{-05} {\rm [M/H]} +0.745832\times 10^{-04} g $ \\
    & $ -0.229508\times 10^{-04} g^2 -0.663820\times 10^{-05}{\rm [M/H]} g $ \\
Si  & $  0.143162\times 10^{-04} +0.342175\times 10^{-04} {\rm [M/H]} -0.102791\times 10^{-04} g $ \\
    & $ -0.119724\times 10^{-05} g^2 -0.174570\times 10^{-04}{\rm [M/H]} g $ \\
Fe  & $  0.169813\times 10^{-03} +0.126582\times 10^{-03} {\rm [M/H]} +0.623935\times 10^{-04} g $ \\
    & $ -0.222898\times 10^{-04} g^2 -0.136709\times 10^{-04}{\rm [M/H]} g $ \\
		\hline
		
	\end{tabular}
\end{table}

This scheme can be applied to any isochrone set to produce approximations for changes wrought by elemental abundance shifts. Note that one would need to locate the subgiant branch separately in order to make the switch between fitting functions. Note also that the temperature shifts on the main sequence are an order of magnitude larger than those inferred from the constant luminosity slices of the RGB. This dichotomy is not echoed in previous work on the effects of varying individual elements \citep{2007ApJ...666..403D,2012ApJ...755...15V}. Our temperature shifts for C, N, O, Si, and Fe for dwarfs are about the same as previous authors, though our Mg shift is less. But the effect we measure on the RGB is smaller than previous authors. We return to this point in the discussion.


\subsection{Integrated light models}

We now briefly recap the state of the integrated light models used for this paper. We compile a set of evolution using ingredients from \citet{2008A&A...482..883M,2009A&A...503..913A,2009A&A...508..355B} with some improvements provided by those authors post-publication. We also include brown dwarfs and stars with $0.06 < m/M_\odot < 0.15$ from our own calculations using MESA. For this paper, we adopt a \citet{2001MNRAS.322..231K} initial mass function.

Altered abundance ratios were included at the isochrone level using the fitting functions of Tables \ref{tab:f} and \ref{tab:f2} and Equation \ref{eq:fit}. They were applied with a judicious cosine bell ``fade through zero'' on the subgiant branch. A principle of superposition was assumed: the $dT(X)$ values for each $X$ are algebraically added when several elements are changed at once. This superposition principle cannot be true to infinite precision, but uncertainties arising from it are negligible compared to other sources of uncertainty.

At the stellar library level, spectral variations that arise from changes in [X/R] were included using the results of line-blanketed synthetic spectra that includes molecules \citep{2009ApJ...694..902L}. Equivalent width style indices were measured from the synthetic spectra, then applied differentially atop recently-refit index fitting functions \citep{Worthey_2014}. For that effort, we merged four empirical stellar libraries to provide absorption feature index information \citep{2006MNRAS.371..703S,1994ApJS...94..687W,2004ApJS..152..251V,2013A&A...549A.129C}. These libraries were smoothed to a common 200 km s$^{-1}$ resolution, and then overlapping stars between them were used to remove residual trends (of which there were almost none). Except for fixing some errors, we left the atmospheric parameters ($T_{\rm eff}$, log $g$, and [Fe/H]) at their literature values and performed new multivariate polynomial fits on an 80-index system \citep{2015MNRAS.453.4431T} that includes the original Lick definitions (but is no longer the ``Lick system.'' We call it simply the 200 km~s$^{-1}$ system.) When corrections to other velocity dispersions are required, we utilize synthetic spectra to generate them.

\begin{figure}
	\includegraphics[width=\columnwidth]{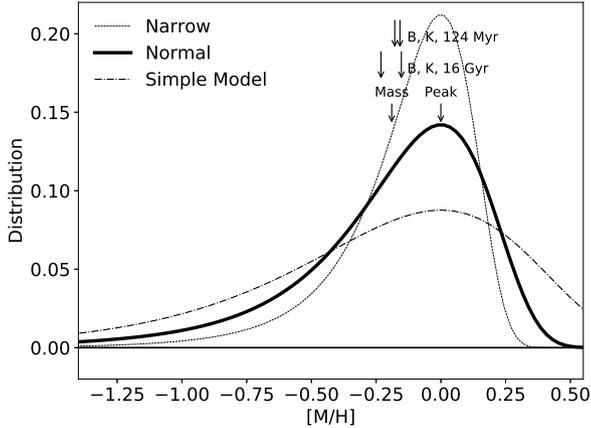}  
    \caption{We adopt the "normal" ADF (bold line) in this paper. For reference, the narrow ADF (thin dotted) and the theoretical simple model ADF (thin dot-dash) are shown. The peak of the distribution is different than the average, since the distribution is asymmetric. The peak, the mass-averaged mean, and $B$ and $K$ flux weighted averages are indicated with labeled arrows.}
    \label{fig:adf}
\end{figure}

After generating simple stellar populations at a grid of ages, metallicities, element changes, and IMF parameters (the latter not exploited in this paper), composite stellar populations were constructed. A single age is retained, but a metallicity spread is included. We use the ``normal'' abundance distribution function (ADF) from \citet{2014MNRAS.445.1538T}, which is considerably more narrow than the simple one zone constant yield chemical evolution model, illustrated in Figure \ref{fig:adf} The metallicity label becomes therefore the peak metallicity of the ADF, but both metal-poor and metal-rich tails are included.

\section{Data sets and Analysis}

We take the non-LINER early-type galaxy averages from SDSS spectra of \cite{2007ApJ...671..243G} as a suitable starting data set. These averages are binned by velocity dispersion ($\sigma = $ 95, 127, 152 175, 205, and 260 km s$^{-1}$). We then measure absorption feature indices and apply the Balmer emission-line corrections of \cite{2010AJ....140..152S} as a function of Mg $b$ strength. All spectra were smoothed to a common velocity dispersion of 300 km s$^{-1}$ for ease of analysis. Synthetic spectral output of the models at high resolution is smoothed to generate velocity dispersion corrections from the native 200 km s$^{-1}$ resolution of the index system to a 300 km s$^{-1}$ smoothing that matches the galaxies.

We constructed an inversion program called COMPFIT (to FIT COMPosite stellar population models) that is an improved version of that used by \cite{2014MNRAS.445.1538T}. It uses hand-picked age- and metallicity-sensitive indices to derive an average (single) age and peak metallicity. The peak metallicity represents the maximum of the asymmetric ADF, and is somewhat higher in value than the average metallicity of the ADF (Fig. \ref{fig:adf}). Because of limitations in the parameter space of the underlying isochrone grid, the peak metallicity is constrained to lie in the range $-0.6 < $ [R/H] $ < 0.4$, where R stands for a generic heavy element. In this paper, with the exception of $\S$3.5, Fe is left in lockstep with the other heavy elements, so that [Fe/R] = [Fe/H] - [R/H] happens to be always zero, and [Fe/H] = [R/H] are conveniently interchangeable in these pages (except $\S$3.5). Because we wish to remain tied to the Fe-peak elements, the hand-picked indices are all either Balmer features for age-sensitivity (H$\delta_F$, H$\gamma_A$, H$\gamma_F$, H$\beta$, and H$\beta_0$) or Fe-sensitive indices (Fe4383, Fe4531, Fe5270, Fe5335, Fe5406, Fe5709).

In a prescribed order, COMPFIT examines various features that are diagnostic of the abundance pattern, weighed by their own sensitivity. With a trial abundance mixture in hand, the program then iterates, improving the age, metallicity, and mixture at each pass until the model matches each set of galaxy indices to an acceptable level. The "prescribed order" maximizes efficiency (compared to relatively blind techniques such as Monte Carlo, Markov Chain Monte Carlo, or Particle Swarm Optimization) by starting with unambiguous elements like Na, which affects primarily the Na D index, and proceeds to elements that affect many indices. Convergence to the point where grid discretization effects emerge is rapid.

Furthermore, because the program is relative speedy, Monte Carlo trials were used to assess overall error and covariances in the output quantities. In such trials, the input measured galaxy indices were varied within their published errors at random. COMPFIT is run again on the scrambled version of the observations. This procedure is repeated until enough trials have been analyzed to provide the covariances and error estimates. In appearance, the Monte Carlo results look like those of Fig. \ref{fig:scat}. Most output quantities show covariances with small amplitude, but age and metallicity remain the most profoundly connected. Note that we did not include IMF in our analysis, and we might have to alter the previous sentence if we had. Although we illustrate the first major breakdown of it below with the case of oxygen, \cite{1994ApJS...95..107W}'s ``theorem of sensitivity'' for abundance ratios holds largely true: abundance ratio measurements are nearly orthogonal to the age-metallicity degeneracy.

\subsection{A conceptual example}

The Sloan averages provide 60 spectral indices for analysis, and we have 24 chemical ratios that can vary, 6 of them carried through to isochrone-level temperature shifts. In addition, ADF peak and single-burst age must vary. We suppressed variation in ADF width, IMF slope, IMF type, IMF low-mass cutoff, and other effects we intend to explore in future papers. 

\begin{figure}
	\includegraphics[width=\columnwidth]{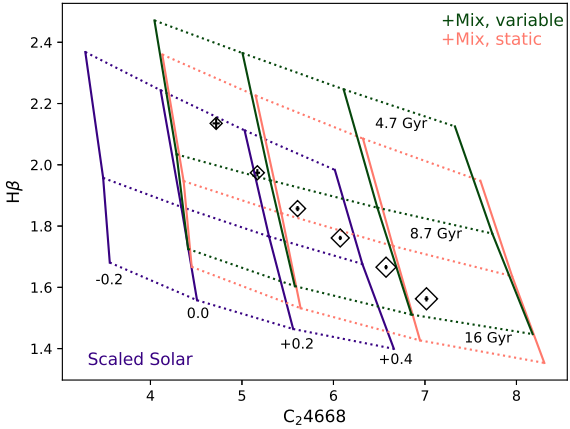}  
    \caption{Models and galaxies in the H$\beta$, C$_2$4668 plane. A scaled solar model grid (blue) compares with a grid with spectral effects included (pink) and a third grid with both spectral effects and isochrone temperature shift included (green). ADF peak metallicities and single-burst ages are as labelled, and Sloan average galaxies are overlaid, with error bars marked inside diamond symbols which increase in size with velocity dispersion.}
    \label{fig:hbc2}
\end{figure}

With such a plethora of variables, machine solution is preferred. However, the logic can be illustrated by looking at a single index-index plane. In Fig. \ref{fig:hbc2}, composite models with ADF peaks at $\{ -0.2, 0.0, 0.2, 0.4\}$ and single-burst ages of $\{ 4.7, 8.7, 16.0\}$ Gyr are shown with Sloan average galaxies for comparison. The chemical mix is chosen to approximately mimic the 260 km s$^{-1}$ bin: [Mg/R] = 0.055, [C/R] = [N/R] [Na/R] = 0.1, and [O/R] = [Si/R] = 0.20. All other ratios are kept at scaled solar.

Firstly and obviously, altered chemical ratios are required for an adequate model-observation match. Secondly, with isochrone temperatures allowed to vary with individual element ratios (dark green grid in the figure), model H$\beta$ strengthens, this time because the MSTO moved hotter. For the purposes of age determination, therefore, inclusion of this new effect is crucial.

Searching for an optimum chemical mixture could be attempted by trial and error among the available index-index planes, roughly $(N^2 - N)/2 = 1770$ planes, given 60 indices. But only one galaxy can be analyzed at a time because chemical mixture strongly couples to mean age.

\subsection{Trends in the root mean square statistic}

Although the figure of merit used in the program is multidimensional, the program can be made to output an RMS statistic for modeled minus observed indices. The RMS is useful for a quick look. As a function of velocity dispersion, the goodness of fit deteriorates about a factor of two from the 95 km s$^{-1}$ bin to the 260 km s$^{-1}$ bin. However, this is in large part due to the fact that there are fewer galaxies to average in the dim bin. The RMS is computed in units of the observational error, so the larger errors in the 95 km~s$^{-1}$ bin drive the appearance of better fit. If we change units on the RMS to be in the native units of the indices, either magnitudes or \AA ngstroms of equivalent width, we lose fidelity because of the varying amplitudes of the indices but we see the same trend, but much reduced. The 260 km s$^{-1}$ bin is fit about 35\% worse than the 95 km s$^{-1}$ bin. This is doubtless due to modeler's difficulties, where we drift from near-solar abundances and its many calibrating stars to super-solar and altered-mixture abundances, where we have essentially zero calibrators.

When the ADF is narrowed by a factor of two from its default width, the models fit better by a few percent for the 95 km~s$^{-1}$ bin and a few percent worse for the 260 km s$^{-1}$ bin. This is weak evidence that the trend we see from resolved-star work in local galaxies \citep{2015A&A...575A..72B,2004PASP..116..295W,1996AJ....112.1975G,2005ApJ...631..820W} is universal. That is, even after averaging over hundreds of SDSS galaxies, the observation that low mass galaxies have a narrower ADF than high mass galaxies seems to remain plausible.

When the flexibility of chemistry is turned off, and we are thrown back to the days of scaled-solar abundances only, even the 95 km s$^{-1}$ bin suffers, with an RMS about 30\% worse than if the spectrum is allowed to change with chemistry. The RMS is 80\% worse for the 260 km s$^{-1}$ bin. This amounts to vindication that our efforts to model nonsolar abundance ratios are worthwhile.

More subtly, if we turn off the isochrone temperature effects from changing chemistry (set $dT(X) = 0$ in equation 1), but keep the spectral effects, the fit is about 1\% worse for all galaxies. The effect we measure with this exercise amounts to mild temperature stretches between dwarfs and giants. This effect should be highly degenerate with age and metallicity and so it is quite interesting that the fitting procedure slightly prefers dynamic isochrone temperatures. 

\subsection{Oxygen age degeneracy}

\begin{figure}
	\includegraphics[width=\columnwidth]{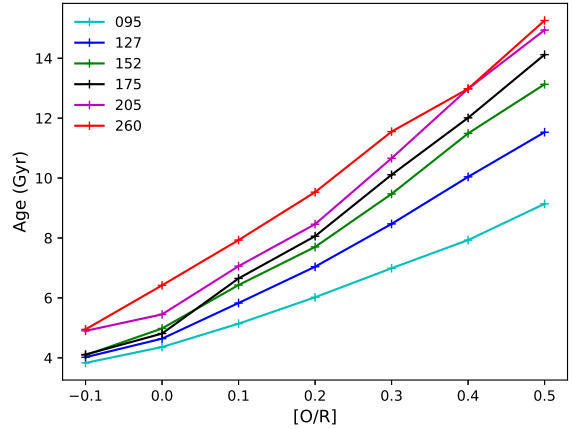}  
    \caption{Inferred ages for the Sloan averages as a function of oxygen abundance. The Sloan averages are binned by velocity dispersion (coloured lines, according to key). The effect of assumed oxygen abundance on age is dramatic.}
    \label{fig:oxy}
\end{figure}

Because the oxygen abundance manifests strongly at the MSTO, we can expect the inferred age to change as a function of [O/R] \citep{2007ApJS..171..146S}. To probe this behavior, we ran a series of inversions with C, N, Na, and Mg left free. Oxygen was held fixed at 0.1-dex increments, and the Sloan averages were inverted under those assumptions. The results  of this experiment are shown in Fig. \ref{fig:oxy}

The change in inferred age is about 2 Gyr per 0.1 dex in [O/R]. Roughly, a factor of two increase in [O/R] leads to nearly a factor of two increase in the inferred age. Because an increase in [O/R] makes the main sequence hotter, an O-enhanced model will appear to be younger than its age. When applying this to a galaxy at fixed line strength, the O-enhanced model must become cooler to compensate, and one convenient way it can make itself cooler is to become older. So with all other variables held constant, we expect a correlation between [O/R] and age. Fig. \ref{fig:oxy} shows that such a correlation exists, and is quite strong even when many variables are left free to potentially compensate for the effect (such as ADF-peak metallicity, [C/R], [N/R], [Mg/R], and [Na/R] --- [Fe/R] and elements not mentioned were held in lockstep with overall metallicity for this experiment). Because O is the most volatile of the elements, however, it dominates, and none of the other elements can, individually, negate its effects.

\begin{figure}
	\includegraphics[width=\columnwidth]{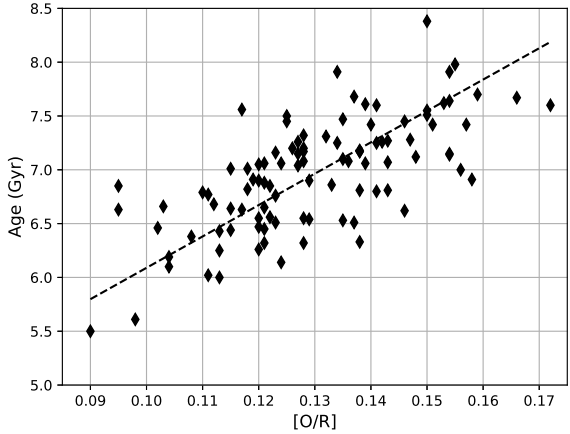}  
    \caption{Realizations of 100 unconstrained model inversions with index data altered randomly within Gaussian error envelopes of magnitude half those published. Age versus [O/R] are shown. The line is a two-error fit with slope $\Delta {\rm age} / \Delta {\rm [O/R]} = 2.9$ Gyr per 0.1 dex interval, somewhat more age-volatile than the $\sim$1.8 slope one would infer from Fig. \ref{fig:oxy}. }
    \label{fig:scat}
\end{figure}

The age-oxygen degeneracy can be seen in another way, as illustrated in Figure \ref{fig:scat}. The figure illustrates the results of Monte Carlo tests wherein artificial errors were introduced into the input catalog of galaxy spectral indices. The index shifts were randomized within a Gaussian probability envelope with $\sigma = 0.5\sigma_{obs}$. Each of 100 such modified index sets were inverted for the Sloan 175 galaxy, and the results in the age - oxygen plane collected into Fig. \ref{fig:scat}.

The line in Fig \ref{fig:scat} was generated using an orthogonal distance regression, and has a slope closer to $\sim$3 Gyr per 0.1 dex in [O/R] rather than $\sim$2 Gyr per dex from the more constrained experiment of Fig. \ref{fig:oxy}. Clearly, the age we derive is heavily dependent upon the oxygen abundance.

\subsection{Magnesium amplification and other effects}

The original worry regarding the enhancement of individual heavy elements rested on an argument based on the H$^-$ opacity in cool stars: a metallic element would have a low ionization potential and hence would donate free electrons in the atmospheres of red giants. The increased H$^-$ opacity would then tend to increase the radius of the star and make it cooler. Thus, increasing the abundance of, for example, Mg, would make the star a little bit cooler. This would tend to (1) strengthen all metallic lines solely due to temperature effects, but also (2) strengthen Mg lines in particular because there would be more absorbers. So Mg features would be doubly enhanced. This positive feedback we termed ``magnesium amplification.'' The phrase rolls off the tongue easily, so it became our jargon for \textit{any} such positive feedback.

The modeling we present finds only minor effects on the giant branch, but substantial effects near the MSTO. So our initial motivation has evaporated, but the concept is still valid. Since the light elements C, N, and O tend to heat up the MSTO while the intermediate-weight ones Si, Mg, and Fe tend to cool it down, we have Si, Mg, or Fe amplification but C, N, or O attenuation. Adding C, N, or O will increase the number of absorbers but also heat the star, weakening metallic feature strengths. This is even more true since we infer C, N, and O abundance primarily from molecular features. But each element tested cooled the giant branch a small amount. Since integrated light also depends on the relative contribution from dwarfs versus giants, the outcome becomes hard to predict without proper modeling.

It is worth noting at this point that heavy element mass fraction $Z$ is not conserved in the preceding. It is correct to say that [R/H] is held constant, but we do simply add Mg (for example) to the mixture, not add Mg and rescale to constant $Z$.

Relative to galaxies, the \textit{signature} of an effect like ``magnesium amplification'' is clear. Models that include it should see greater predicted ranges in absorption feature strengths compared to models that do not include it. The signature for inferred properties of galaxies is therefore that they should appear less extreme (for inferred quantities such as [Mg/Fe]) with the new models.

With those remarks as preface, let us examine results for some observations we have examined before \citep{Worthey_2014}, averages for non-LINER early-type galaxies taken from the SDSS by \cite{2007ApJ...671..243G}, the treatment of which was described at the beginning of the section. During inversion, most elements were held in lockstep with Fe. C, N, O, Mg, and Na were allowed to vary without constraint. The special case of Si, which does not strongly affect line strengths, was forced to a plausible variable pattern with velocity dispersion \citep{2014ApJ...780...33C,Worthey_2014}, such that [Si/R] = 0.0 for the 95 km s$^{-1}$ bin, and increases by 0.04 dex each bin for a maximum of [Si/R] = 0.2 for the 260 km s$^{-1}$ bin.

\begin{figure}
	\includegraphics[width=\columnwidth]{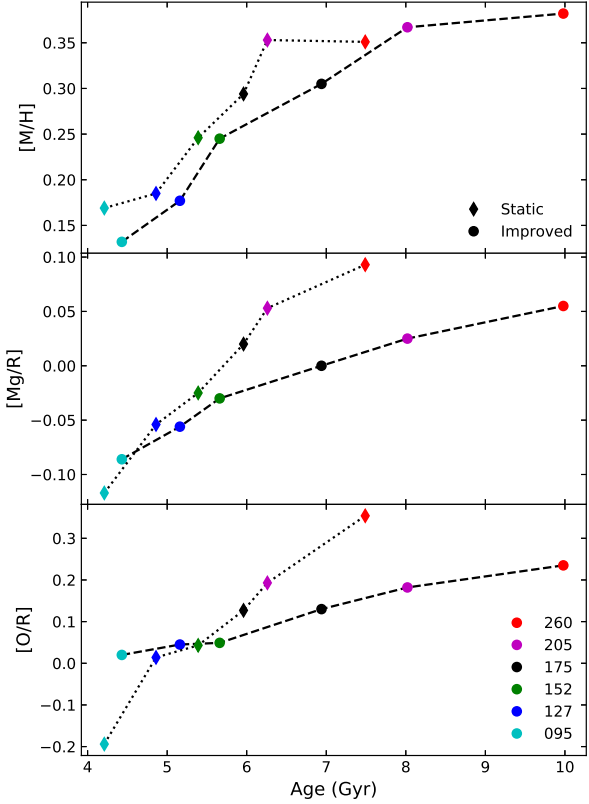}  
    \caption{Shift in results when isochrone temperatures go from static to variable. ADF peak metallicity [M/H], [Mg/R], and [O/R] are plotted as a function of mean CSP age. These are model inversions. The improved, variable-temperature case (circles) shows a larger span of ages but a smaller range of abundance parameters than the static-temperature case (diamonds). Velocity dispersion bins for the Sloan averages are colour coded as indicated. }
    \label{fig:mg1}
\end{figure}

\begin{figure}
	\includegraphics[width=\columnwidth]{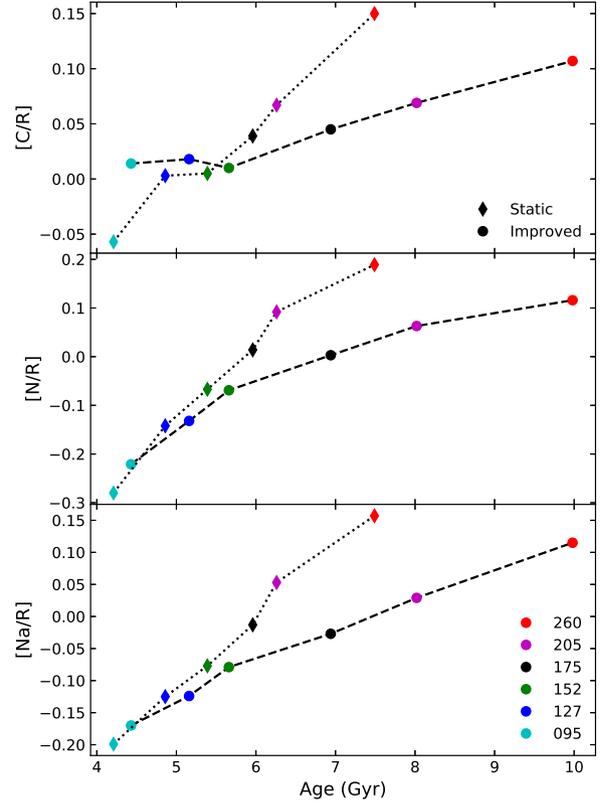}  
    \caption{Shift in results when isochrone temperatures go from static to variable. [C/R], [N/R], and [Na/R] are plotted as a function of mean CSP age. These are model inversions. The improved, variable-temperature case (circles) shows a larger span of ages but a smaller range of abundance parameters than the static-temperature case (diamonds). Velocity dispersion bins for the Sloan averages are colour coded as indicated. }
    \label{fig:mg2}
\end{figure}

Figures \ref{fig:mg1} and \ref{fig:mg2} plot abundance results versus age. The signature of ``magnesium amplification,'' a decrease in range of inferred abundance quantities is apparent in all the elements. Interestingly, it is even apparent in sodium. Sodium is a minor element that is not expected to affect isochrone temperature (and in any case does not, here, since we did not model its effects). However it manages to also behave in muted fashion with the new temperature-variable isochrones compared with the static isochrone models.

In contrast to the element ratios, the range of mean ages increased. Also, the ages are older, overall. We did not predict this outcome. The most we could have predicted is that \textit{if} the [O/R] range decreased, then the age range should therefore increase, due to the oxygen-age degeneracy. However, the new models clearly favor older ages, especially for the galaxies with higher velocity dispersions and more extreme abundance ratios.

\begin{figure}
	\includegraphics[width=\columnwidth]{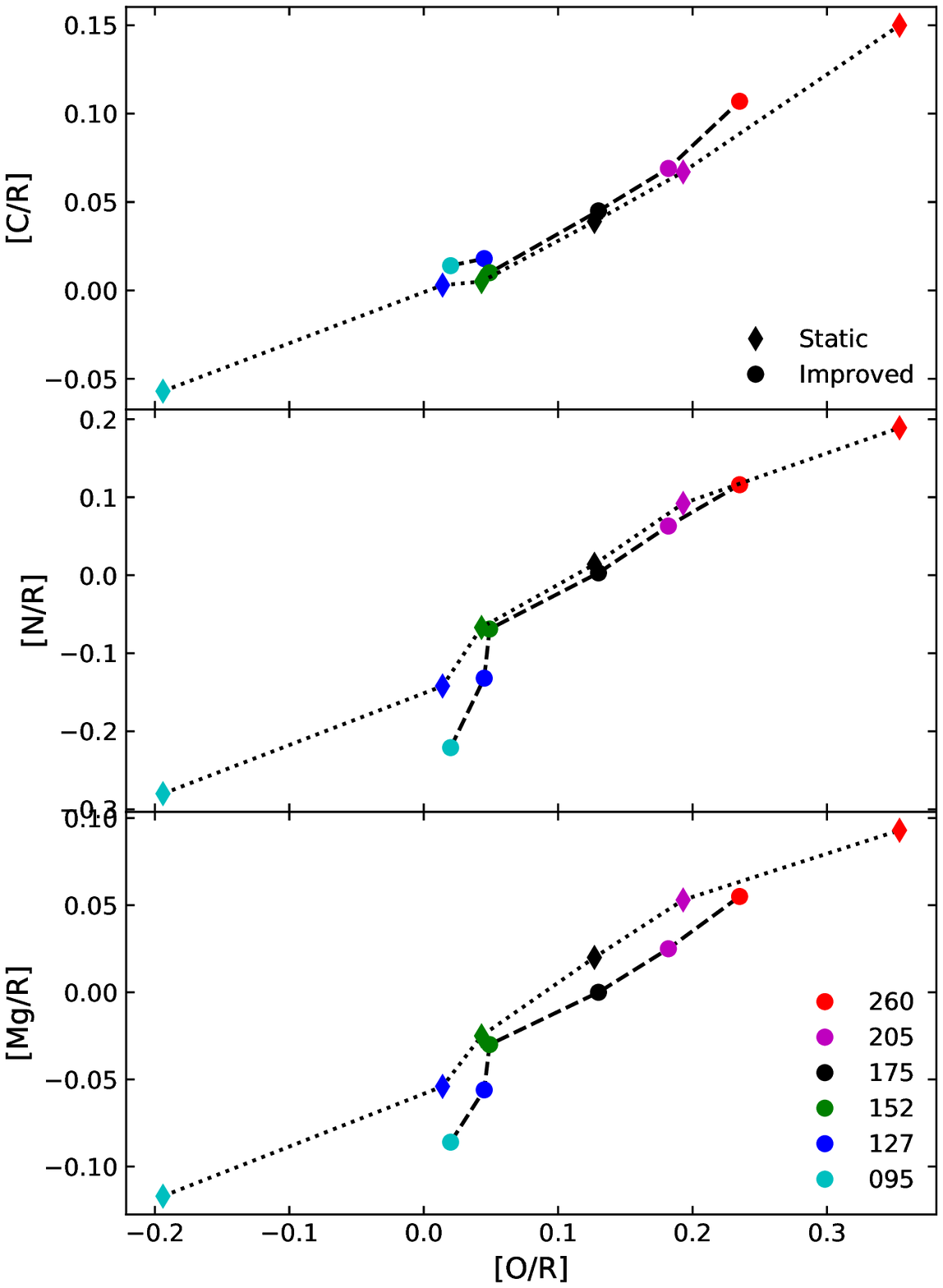}  
    \caption{Shift in results when isochrone temperatures go from static to variable. The span of abundance parameters, [C/R], [N/R], [Mg/R], and [O/R] all decrease in range when the isochrone temperatures go from fixed (diamonds) to variable (circles), but their trends and ratios stay approximately the same.}
    \label{fig:mg3}
\end{figure}

The relative trends among light elements are shown in Fig. \ref{fig:mg3}. In these cases, there is little change in the interrelated behavior of the various abundance ratios whether the isochrones are static or temperature-variable. The range of variation decreased for all elements, but neither slope nor zeropoint shifted by much.

\subsection{An Fe-peak experiment}

If oxygen is the most effective heater (Figs \ref{fig:eep_ms} \& \ref{fig:eep_rgb}) then iron is the most effective cooler. If including variable oxygen led to an increased age span, would including variable Fe lead to a decreased age span? In all the preceding, the Fe-peak elements have been held in lockstep with the rest of the heavy elements. In relaxing this restriction, we sail into somewhat perilous waters. With the extra freedom, the inversion program may converge less well or even go unstable, since it operates unconstrained. Also, the abundance mixture used for computing the isochrones resembles the output mixture less and less. And finally, we are less than convinced that the temperature dependence for Fe is correct on the giant branch (see discussion).

\begin{figure}
	\includegraphics[width=\columnwidth]{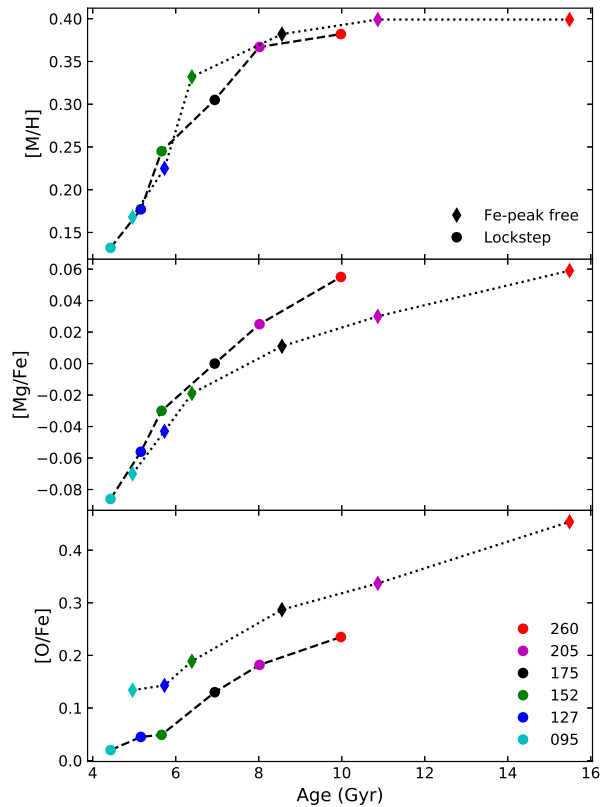}  
    \caption{Shift in results when the Fe-peak is allowed to vary along with other elements, isochrone temperatures variable in both cases. ADF peak metallicity [M/H], [Mg/Fe], and [O/Fe] are plotted as a function of mean CSP age. These are model inversions. The variable Fe-peak case (diamonds) shows an even larger span of ages than the lockstep Fe-peak case (circles). Velocity dispersion bins for the Sloan averages are colour coded as indicated. }
    \label{fig:fe1}
\end{figure}

\begin{figure}
	\includegraphics[width=\columnwidth]{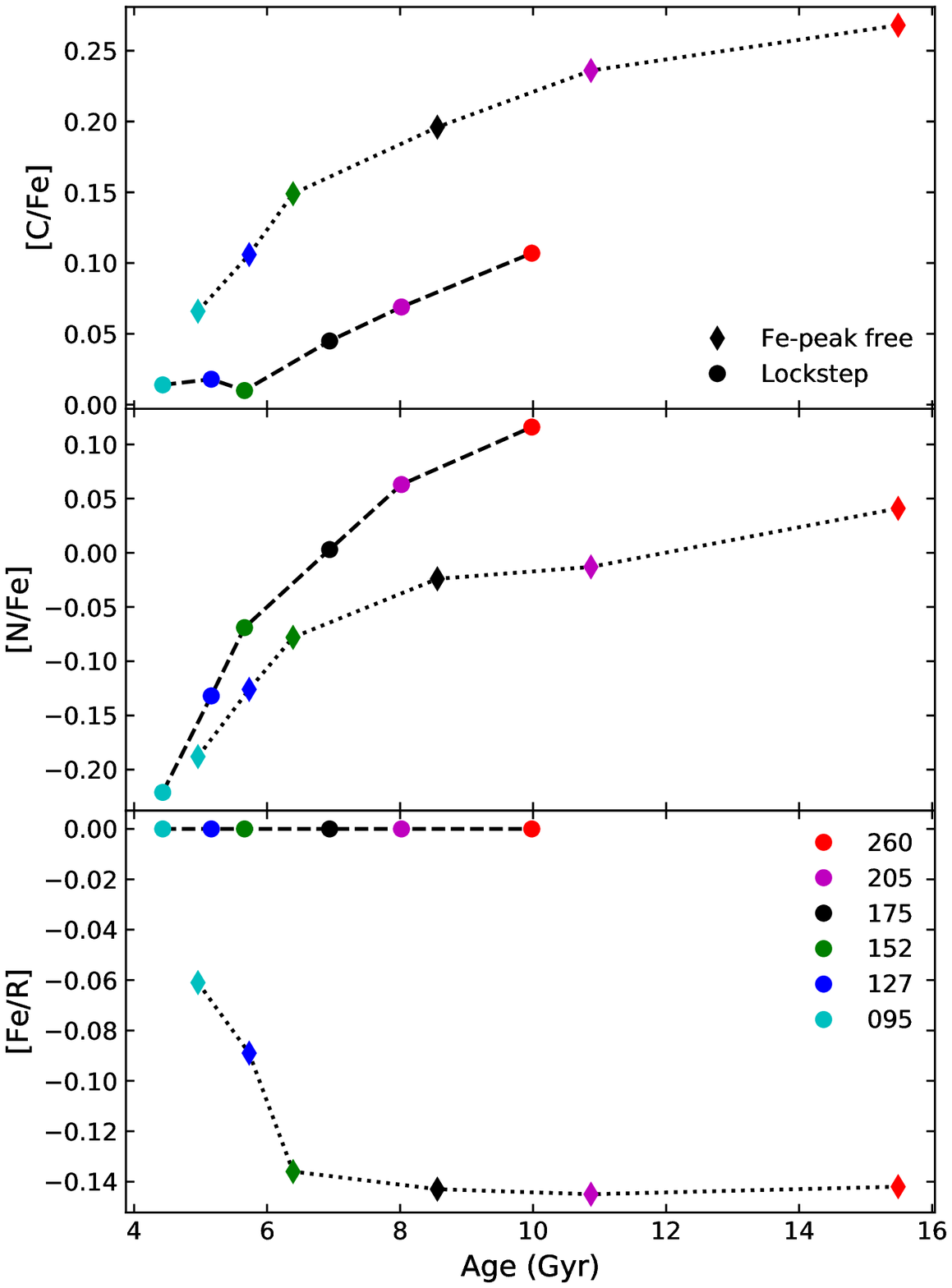}  
    \caption{Shift in results when the Fe-peak is allowed to vary along with other elements, isochrone temperatures variable in both cases. ADF peak metallicity [C/Fe], [N/Fe], and [Fe/R] are plotted as a function of mean CSP age. These are model inversions. The variable Fe-peak case (diamonds) shows an even larger span of ages than the lockstep Fe-peak case (circles). Velocity dispersion bins for the Sloan averages are colour coded as indicated. }
    \label{fig:fe2}
\end{figure}

Nevertheless, we locked Fe together with Cr, Mn, Co, and Ni and ran inversions. No runaway convergences or wild values resulted. Highlights from this experiment are shown in Figs. \ref{fig:fe1} and \ref{fig:fe2}. We see in the lowest panel of Fig. \ref{fig:fe2} that the Fe-peak drops to [Fe/R] $\approx -0.14$. This should drive the lighter elements to higher values, and should also drive the overall heavy element content to higher values. In fact, for the two largest velocity dispersion bins (Fig. \ref{fig:fe1}, top) we run into the top end of the peak metallicity range at +0.4. These two bins should be largely ignored, therefore. 

Including Fe increased O and C, while dropping N and Mg. Na showed little change, and is not plotted. Inferred ages increased even more, as did the age range. This leads us to conclude that the increase in age range is a robust result, largely driven by oxygen's temperature dependence. 

The experiment also leads to a sense of caution about absolute values of the abundance ratios. Until the temperature dependencies of isochrones are firmly settled, some volatility will remain in measuring abundance ratios.

\section{Discussion}

For the galaxy results, the observed compression of at least some of the ranges of the individual element ratios was an expected outcome. The expansion of the age range was not. However, this stretched age range is a robust consequence of the improved temperature sensitivity in combination with the trends in abundance ratios seen in early-type galaxies as exemplified by the Sloan averages.

The reduced swings in element ratios will have several effects on our understanding of galaxy evolution. The more modest swings in chemistry might be easier for chemical evolution models to reproduce. If the [Mg/Fe] trend is due to star formation timescale, as a lot of authors assume \citep{2002Ap&SS.281..371T,1994A&A...288...57M,2006MNRAS.372..265M,2009A&A...499..409C}, then our span (c.f. Fig \ref{fig:mg1}) of 
$\Delta \mathrm{[Mg/R]}\approx 0.1$ 
should be less stringent than the previous $\approx 0.3$ \citep{2000AJ....120..165T}. The same argument applies to hypotheses involving high-mass IMF modulation or binary fraction changes as function of progenitor galaxy mass. All such scenarios can be less extreme. 

Abundance ratio gradients also, although weak on average already,  \citep{1999PASP..111..919H,2015ApJ...808...26R} will flatten even more on this newer, more accurate scale. \cite{1992ApJ...398...69W} pointed out that flat gradients in abundance ratios imply galaxy-global enhancements in light elements, which is surely an important constraint when deciding among hypothetical mechanisms for light element enhancement.
In particular, the reduced chemical spread might indicate a vigorous level of gas-poor merging because such "dry" mergers tend to erase abundance gradients and chemically homogenize the pre-merger fragments \citep{2009A&A...499..427D}.

We note that the element with the largest dynamic range is now nitrogen (c.f. Fig \ref{fig:mg2}). The nucleosynthetic origin of this element includes a robust and possibly dominant contribution from intermediate-mass stars \citep{2003MNRAS.339...63C}. But there is a nitrogen-sigma relation, too \citep{1998PASP..110..888W}, which tosses a complicating factor into the debate over the origin of excess light elements (upper IMF strength, star formation timescales, wind truncations, etc.).

The Mg-$\sigma$ relation \citep{1993ApJ...411..153B,2003ApJ...586...17W} would now appear in light of our results to have a significant tie to mean age, as well as to Mg abundance itself. This could still be mostly illusion, however, as the galaxy bins average over many galaxies. Previous studies \citep{1998PASP..110..888W,2007ApJS..171..146S} indicate that smaller galaxies have a greater age scatter than the very large ones at high velocity dispersion (and that galaxy selection has great influence). The SDSS data alone cannot distinguish between size-age and size-scatter relations.

Generally, our age results are consistent with galaxy-counting at large lookback times. Redshift surveys beyond $z\approx 1.5$ show luminosity function growth over cosmic time \citep{2004ApJ...608..752B,2006ApJ...651..120B,2007ApJ...665..265F,2010ApJ...709..644I,2010ApJ...718.1158L,2014ApJ...783...85T}. When split by colour the red sequence can be loosely equated to early type galaxies. Roughly, these galaxies have tripled in number since $z\approx 1.5$ and doubled in number since $z=1.0$, although a modest red sequence resembling today's spheroids appears to have been in place by $z\approx2$ \citep{2013ApJ...774...47L}, at the peak of the universe's star formation \citep{2014ARA&A..52..415M}. 

Redshift $z\approx 1.8$ corresponds to a lookback time of 10 Gyr, the age of the most massive bin of our SDSS sample, and therefore the most extreme early universe overdensities. This age harmonizes with high redshift results. The mean age-mass correlation, often termed downsizing \citep{2006MNRAS.372..933N}, sharpens with the increased age range that we show. Notably, gas-poor mergers tend to erase previously existing stellar abundance gradients \citep{2009A&A...499..427D} and also tend to feature in scenarios of red-sequence galaxy mass buildup \citep{2017MNRAS.467.3083R}. Our results seem to favor a healthy number of "dry" mergers during galaxy formation.

We noted in $\S2.1$ that our calculated temperature shift on the RGB as a function of individual element with $Z$ held fixed were smaller than other authors have calculated using different codes and assumptions. Our outer boundary condition is a gray model atmosphere. The mean molecular weight and mean opacity vary as a function of individual element mixture, but this may not be adequate to capture the essential physics. This is an item that should be improved for future grid calculations, but of course is also an item fraught with considerable uncertainty.

We can predict what would happen to the results of this paper if the RGB temperatures varied more. Our oxygen dependence seems secure in comparison with previous calculations, so the increased age range of the SDSS galaxies would continue to be a firm conclusion. The inferred ranges of the abundances, however, would shrink even more, especially that of Mg, because the temperature throw we predict for Mg is smaller than \citet{2007ApJ...666..403D}, even for dwarfs. In other words, magnesium amplification would be more effective.

We did not explore the possible age effects of helium in this work because of the difficulty (or, perhaps, impossibility) of measuring the abundance of this element from an integrated light absorption line spectrum. Knowledge of plausible helium mixtures must come from other lines of reasoning, such as emission line studies or studies of the UV upturn phenomenon \citep{2018MNRAS.476.1010A}. 

Future model improvements should include a critical exploration of the effects of detailed abundance changes on the outer boundary conditions of stellar models. Too, our simple formulae for temperature shifts, while useful, are an incomplete description. The effects on stellar lifetimes and luminosities should also be included. However, the effects of such considerations as binary evolution effects and blue straggler creation will have a larger spectra effect, so they should be address first.  

As regards applications, the present CSP models could be employed well to individual galaxies near and far \citep{2016Natur.540..248K}. To gauge the scatter in the abundance and age properties might tell us more about galaxy assembly than measuring grand averages. Spatially resolved galaxy observations would reveal radial profiles in abundance ratios, and these may not be smooth, but may display substructure that could then be used to decipher galaxy histories on an individual basis \citep{2001AJ....121..244S,1996AJ....111.1512S,1992A&A...258..250B,2021A&A...650A..50E,2021A&A...647A.181O,2019MNRAS.485.3215T}. 

Abundance ratios should be taken into account when examining colour-magnitude diagrams of Galactic star clusters, especially the metal rich ones.

\section{Conclusions}

Composite stellar population models are compared with averaged early type galaxy spectra. Inclusion of explicit temperature changes as a function of individual element abundances sparks a cascade of effects.

\begin{itemize}
    \item Over the range of velocity dispersion in early type galaxies (95 $< \sigma <$ 260 km $s^{-1}$) the dynamic ranges of inferred abundance ratios all decrease. [C/R] and [Mg/R] now span only 0.1 dex, [M/H] and [O/R] span 0.2 dex, and [N/R] and [Na/R] span 0.3 dex. 
    \item Over the range of velocity dispersion in early type galaxies the dynamic range of inferred age range increases to 4.5 Gyr for small galaxies and 10 Gyr for the largest galaxies.
    
\end{itemize}

Various experiments with our parameter-inverter program tended to solidify our conclusions, including setting free the Fe-peak abundances to wander away from the base isochrones during inversions.

Previous estimates of abundance ratios such as [Mg/Fe] and [O/Fe] in massive early-type galaxies are overestimates by roughly a factor of 0.1 dex, and shifts occur on the low-mass end as well, in the opposite sense, for an overall shrinking of the inferred range of abundance ratio variation. Galaxy abundance ratio gradients will flatten.  Age may drive the Mg-$\sigma$ relation more than had been previously appreciated. Locally, the temperature shifts caused by abundance ratio effects should be included for comparison of isochrones with star cluster colour magnitude diagrams, because shifts can be significant.

\section{Data availability}

Much of the data underlying this article are publicly available. Any additional data underlying this article will be shared upon reasonable request to the corresponding author.






\bibliographystyle{mnras}
\bibliography{isochrone} 





\bsp	
\label{lastpage}
\end{document}